\documentclass[review]{elsarticle}

\usepackage{hyperref}

\journal{Acta Materialia}

\usepackage{natbib}
\usepackage{amsmath}
\usepackage{amsfonts}
\usepackage{amssymb}
\usepackage{bm}
\usepackage[thinlines,thiklines]{easybmat}
\usepackage{epsfig}

\usepackage{color}

\DeclareTextSymbol{\degre}{T1}{6}
\DeclareTextSymbol{\degre}{OT1}{23}

\newcommand{\vect}[1]  { \bm{\mathrm{#1}} }
\newcommand{\matr}[1]  { \bm{\mathrm{#1}} }

\begin{document}

\begin{frontmatter}

\title{Entangled Single-Wire NiTi Material: \\ a Porous Metal with \\Tunable Superelastic and Shape Memory Properties}

\author[3SR1,3SR2,SIMAP1,SIMAP2]{B. Gadot}
\author[3SR1,3SR2]{O. Riu Martinez}
\author[3SR1,3SR2]{S. Rolland du Roscoat}
\author[SIMAP1,SIMAP2]{D. Bouvard}
\author[ILM]{D. Rodney}
\author[3SR1,3SR2]{L. Org\'eas\corref{mycorrespondingauthor}}
\cortext[mycorrespondingauthor]{Corresponding author}
\ead{Laurent.Orgeas@3sr-grenoble.fr}

\address[3SR1]{CNRS, 3SR Lab, F-38000 Grenoble, France}
\address[3SR2]{Univ. Grenoble Alpes, 3SR Lab, F-38000 Grenoble, France}
\address[SIMAP1]{CNRS, SIMAP, F-38000 Grenoble, France}
\address[SIMAP2]{Univ. Grenoble Alpes, SIMAP, F-38000 Grenoble, France}
\address[ILM]{Institut Lumi\`ere Mati\`ere, Universit\'e Claude Bernard Lyon 1, 
F-69622 Villeurbanne, France}

\begin{abstract}
NiTi porous materials with unprecedented superelasticity and shape memory were manufactured by self-entangling, compacting and heat treating NiTi wires. The versatile processing route used here allows to produce entanglements of either superelastic or ferroelastic wires with tunable mesostructures. Three dimensional (3D) X-ray microtomography shows that the entanglement mesostructure is homogeneous and isotropic. The thermomechanical compressive behavior of the entanglements was studied using optical measurements of the local strain field. At all relative densities investigated here ($\sim 25-40\%$), entanglements with superelastic wires exhibit remarkable macroscale superelasticity, even after compressions up to 25\%, large damping capacity, discrete memory effect and weak strain-rate and temperature dependencies. Entanglements with ferroelastic wires resemble standard elastoplastic fibrous systems with pronounced residual strain after unloading. However, a full recovery is obtained by heating the samples, demonstrating a large shape memory effect at least up to 16\% strain.
\end{abstract}

\begin{keyword}
NiTi porous materials, Entangled Monofilament, Architected Materials 
\end{keyword}

\end{frontmatter}


\section{\label{par:int}Introduction}
Due to their biocompatibility and potentially interesting mechanical properties, porous NiTi materials are promising architected media, in particular for biomedical applications such as bone implants \citep{Bansiddhi08,Geetha09}.  Up to now, most of these materials were processed by powder metallurgy technologies, \textit{e.g.} self-propagating high-temperature synthesis \citep{Biswas05,Jang06}, spark plasma sintering \citep{Zhao05}, hot isostatic pressing \citep{Lagoudas02} and conventional sintering \citep{Bertheville06, Aydogmus09, Aydogmus12}. Many difficulties arise with these processing routes, including the growth of a brittle oxide skin and a difficult control of (i) the porous mesostructure, (ii) the microstructure and (iii) the thermomechanical properties of the NiTi solid phase. Intense research was devoted to overcome these difficulties, largely improving the resulting mesostructures and biocompatibility \citep{Bansiddhi08,Geetha09}. However, the macroscale mechanical properties of NiTi porous media, such as their superelasticity and shape memory, are still lower than in bulk NiTi alloys. Furthermore, porous alloys suffer from large residual strains induced by stress concentrations that arise at grain boundaries and necks and lead to local plastic deformations and ruptures \citep{Itin94}. 

In parallel, there is a growing research activity on the development of new types of metal porous media made of self-entangled monofilaments \citep{Liu08,Tan09,Liu10,Courtois12,He12,Zhang13,Jiang14}. These fibrous materials are often called "entangled metallic wire materials" or "metal rubber" because their mechanical behavior reminds that of elastomers. Compared to NiTi foams, manufacturing is simpler, consisting in cold/hot working treatments to create a significant amount of self-entanglements of the wire. Up to now, various metal wires have been used: steel \citep{Liu08,Liu09,Courtois12}, Al \citep{Tan09}, Ti \citep{Liu10,He12,Jiang14} and Ni \citep{Zhang13}. To strengthen their mechanical response, sintering or glueing operations can be used \citep{Liu08,Liu09,Liu10,He12,Jiang14}. In all cases however, the mechanical behavior still exhibits limited reversible strain, typically $0.01-0.1$ in compression. This is mainly due to the low yield stress of the elastoplastic wires, which deform plastically during the macroscopic deformation of the entanglements. To limit this effect, \citet{Courtois12} entangled high yield stress perlitic steel wires, but the resulting architecture was highly heterogeneous due to the important spring back effect induced by the stiff wire and the entanglements had to be confined inside a solid container, thus severely limiting the use of perlitic steel wires.

In this context, the main objective of the present work is to process a novel homogeneous porous NiTi material that exhibits important superelastic and shape memory effects, \textit{i.e.} with a strain reversibility largely above those observed with other NiTi porous media, as well as other entangled monofilaments. To circumvent the difficulties mentioned above, we revisited and adapted the manufacturing route of single-wire materials to the case of a commercial biomedical grade NiTi alloy. Moreover, tuning the processing parameters allowed us to tailor  the mesostructure, superelasticity behavior and shape memory properties of these new NiTi porous materials. In the following, the methods used to prepare and characterize these materials are presented (Section \ref{par:procedure}), together with both the properties of the shaped wires (Section \ref{par:wire}) and their mesostructures (Section \ref{par:micro}). The macroscale thermomechanical properties of the entanglements are then investigated and discussed (Section \ref{par:resu}).

\section{Experimental procedure}
\label{par:procedure}

\subsection{Material and sample processing}
\label{par:mater}

\begin{figure}
\begin{center}
\includegraphics[width = 6 cm]{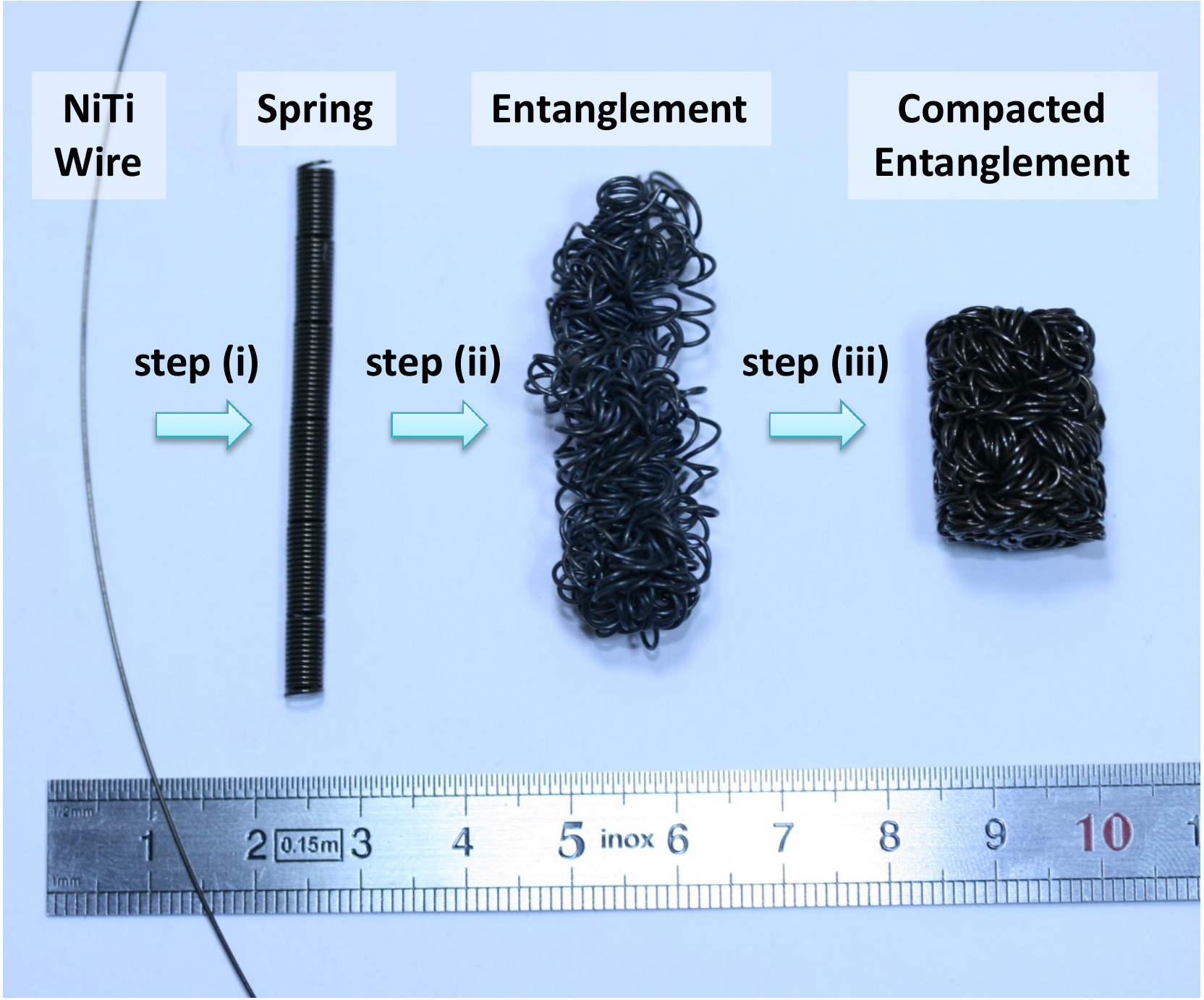} 
\caption{\label{fig_processing} Illustration of the various shape setting operations followed to produce a homogeneous NiTi entangled single-wire material.}
\end {center}
\end {figure}

The route used to manufacture NiTi porous materials is inspired from the work of \citet{Tan09} for ductile and elastoplastic Al alloy wires. As illustrated in Fig.~\ref{fig_processing}, there are three steps to process homogeneous specimens: (i) shaping a spring from a straight wire using plastic deformation, (ii) entangling the spring into a spring ball, and (iii) setting the final shape and consolidating the entanglement by close die compaction.

A standard biomedical-grade Ni-50.8at.\%Ti supplied by Forth Wayne Metals in cold worked state with a diameter $d=$~500~$\mu$m was chosen for its capability to display a superelastic behavior at room temperature and a shape memory effect after suitable short and low-temperature heat treatments. The spring shape in step (i) was achieved by rolling the wire around a 3 mm diameter threaded cylindrical bar, with a pitch close to the wire diameter. A cylindrical steel close die mounted on tension-compression testing machine (MTS DY34, maximum force 100kN) equipped with a furnace was used for step (iii). The final samples have a cylindrical shape with a height $H_0=27$ mm and a diameter $D_0=20$ mm. Samples of various initial relative density, or fiber content, $\bar\phi$, ranging from 0.27 to 0.36, were produced by starting with straight wire of lengths between $11.66$~m and $15.55$~m.

In its initial cold worked state, the NiTi wire is elasto-plastic with a high yield stress ($\approx$~1800~MPa), pronounced strain hardening and a rather small ultimate tensile strain ($\approx$~0.04). These features are not compatible with the low yield stress and high ductility necessary to process homogeneous self-entangled materials, as already pointed out in the Introduction \citep{Stoop08, Stoop11, Courtois12}. To overcome these difficulties, the deformed wires were subjected to short optimised heat treatments \citep{Pelton00,Favier06} at 623~K for 2~min (followed by water quenching) after both steps (i) and (iii).

The temperature and duration of the heat treatments were also chosen to ensure that the wire was superelastic at room temperature \citep{Pelton00,Favier06}, a state called \textit{state 1} hereafter. Therefrom, the entanglement thermomechanical properties could be tailored using additional heat treatments. For example, the superelastic domain could be shifted to higher temperatures by heating the samples at 723~K during 1~h. The resulting new state of the entangled wire, which is now ferroelastic at room temperature, will be called \textit{state 2}. 

\subsection{Mesostructure characterization}
\label{par:tomo}

\begin{figure}
\setlength{\unitlength}{1cm}
\begin{center}
\includegraphics[width = 9.4 cm]{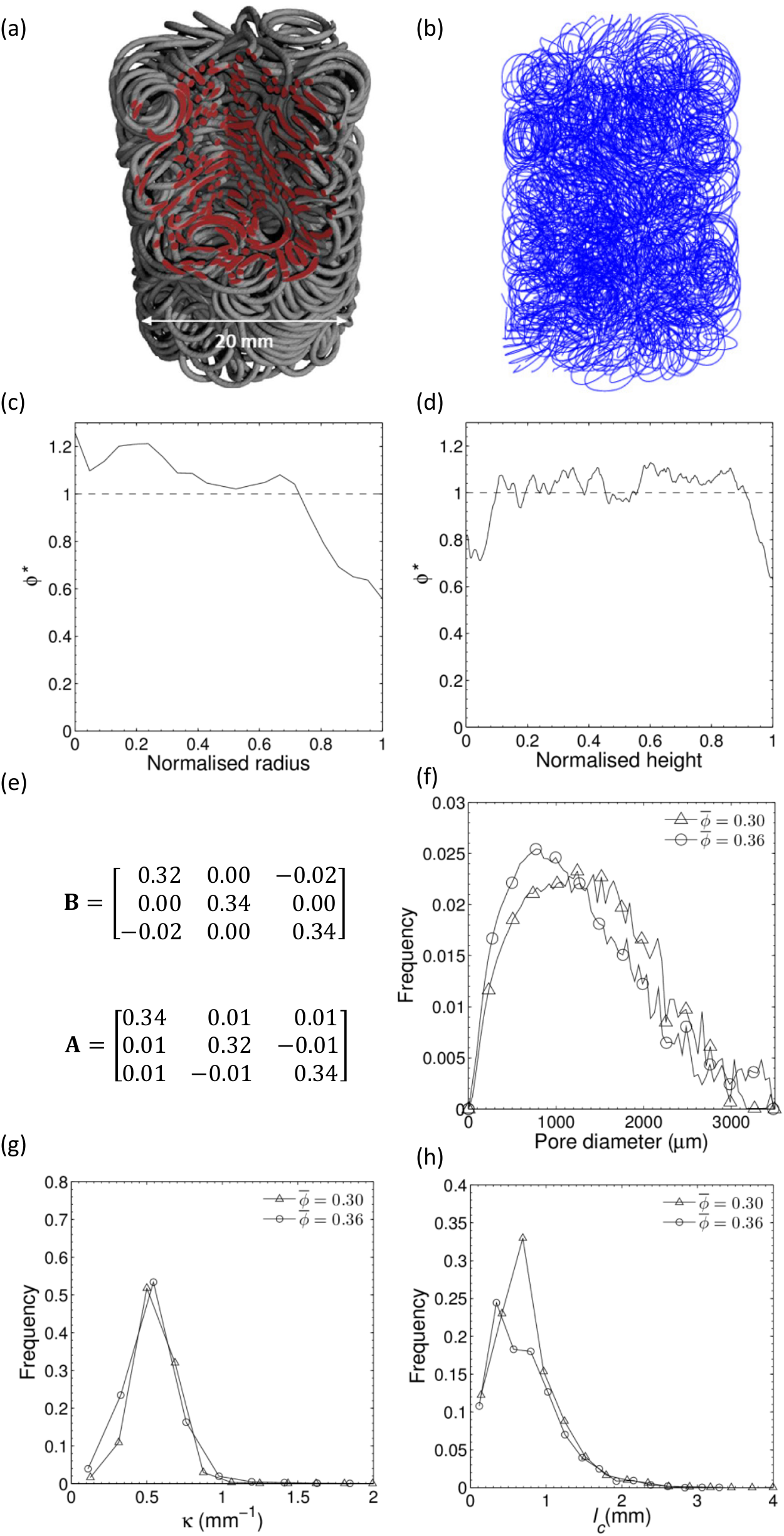} 
\caption{\label{fig_tomo} 3D view (a), centerline (b) and second order fiber and contact orientation tensors $\matr{A}$ and $\matr{B}$ (e) of a NiTi entanglement with a mean fiber content $\bar\phi=$~0.36, together with fiber content profiles $\phi^*=\phi/\bar\phi$ along its radius (c) and height (d). The pore size (f), mean curvatures (g) and contact lengths (h) distributions are also shown for both scanned samples.}
\end{center}
\end{figure}

Information about the entanglement mesostructures was obtained by scanning two typical samples,\textit{ i.e.} with $\bar\phi$= 0.3 and 0.36, inside an X-ray laboratory microtomograph (RX solution, 3SR Lab, Grenoble, France). The scans were obtained with 1200 2D radiographs onto a 1914 $\times$ 1580 pixel flat panel (leading to a voxel size = 23 $\times$ 23 $\times$ 23 $\mu$m$^3$), a scanning intensity of 200~$\mu$A, a tension of 150~kV, and an exposure time per radiograph of 0.25~s (to restrain the noise, an average of six radiographs per 2D image was used). To reduce beam hardening effects, a 1~mm Cu filter was positioned between the X-ray source and the sample. After reconstruction followed by filtering and thresholding operations, 3D binarized images of the samples were obtained. An example is shown in Fig. \ref{fig_tomo}(a). 
The images were used to analyse:

\begin{itemize}
\item the local fiber content profiles along the height and radius of the samples (Fig. \ref{fig_tomo}(c-d)).
\item the pore size distribution, by using morphological operations from a hexagonal structural element with an eight voxel size (Fig. \ref{fig_tomo}(f))  \cite{Maire07,Chalencon10}. 
\item the anisotropy, connectivity and mean curvature of the wire, by following the method proposed by \citet{Latil11} and \citet{Orgeas12}. As illustrated in Fig. \ref{fig_tomo}(b), the centerline of the wire was first detected by (i) estimating the 3D Euclidian distance map of the fibrous phase, (ii) thresholding the resulting grey scale image at an intermediate grey value to eliminate contact zones from the image, (iii) skeletonizing the thinned wire, (iv) smoothing the resulting skeleton to obtain its centreline. The skeleton was discretized in elementary segments. The segment length was chosen equal to $2d$, twice the fiber diameter. Each segment $i$ has a curvilinear abscissa and unit tangent vector,  respectively noted $s_i$ and $\vect{p}_i$. For the two scanned samples, \textit{i.e.} with $\bar\phi$= 0.3 and 0.36, the number of segments $N$ was 12960 and 15552, respectively. 

The mean orientation of the centerline was determined by computing the second order fiber orientation tensor $\matr{A}$ \citep{Advani87} (see Fig.\ref{fig_tomo}(e)):
\begin{equation}
\matr{A}=\frac{1}{N}\sum_{i=1}^N\vect{p}_i\otimes\vect{p}_i.
\end{equation} 
The variation of tangent vector $\vect{p}_i$ along the curvilinear abscissa $s_i$ of the wire was used to compute the local mean curvature $\kappa_i$ (see Fig. \ref{fig_tomo}(g)):
\begin{equation}
\kappa_i=\big|\big|\frac{\mathrm{d}\vect{p}}{\mathrm{d}s}\big|\big|_{s_i}.
\end{equation} 
Self-contacts were defined as pairs of elementary segments with a distance less that the fiber diameter. Each contact $j$ was ascribed an orientation unit vector $\vect{n}_j$ normal to its contact plane. The mean contact orientation was then estimated from the second order contact orientation tensor $\matr{B}$ (see Fig. \ref{fig_tomo}(e)):
\begin{equation}
\matr{B}=\frac{1}{M}\sum_{j=1}^M\vect{n}_j\otimes\vect{n}_j,
\end{equation}
where $M$ is the total number of elementary contacts. We should note that a single contact may be composed of several elementary contacts that link neighboring segments. We thus distinguish between elementary contacts as used in the above equation, whose number depends on the choice of discretization length, and full contacts that were obtained after agglomeration of elementary contacts between neighboring segments. In the following, we will note $n$, the number of full contacts per unit fiber length. Their length $l_{c}$ was also computed and analyzed (see Fig. \ref{fig_tomo}(h)).
\end{itemize}

\subsection{Thermomechanical characterization}
\label{par:thermomeca}

The local stress-free forward and reverse martensitic transformations of the processed entanglements were analyzed by cutting small wire pieces inside samples and subjecting them to DSC measurements. Straight wires that had undergone the same thermal history as the entanglements were also tested. The tensile mechanical behavior was assessed at various temperatures ranging from 293~K to 353~K, subjecting the wires to load-unload cycles up to a maximum tensile strain of $\approx$ 0.08 at a strain rate of 10$^{-3}$s$^{-1}$ (testing machine Gabo Eplexor 500~N equipped with a furnace). Furthermore, the processed entanglements were subjected to compression cycles at various temperatures and strain rates ranging respectively from 293~K to 353~K and from $4~10^{-4}$~s$^{-1}$ to $3.7~10^{-2}$~s$^{-1}$. The samples were placed inside a compression rheometer with parallel plates equipped with an inner heating system \citep{Guiraud12b} and mounted on a mechanical machine (MTS DY26, maximum force 100~kN). During the tests, the recorded compression force, $F$, and the current sample height, $H$, were used to estimate the nominal compression stress $\sigma=4F/\pi D_0^2$ and global axial compression Hencky strain $\varepsilon_g=\ln (H/H_0)$. In the following, the reference testing temperature and strain rate are set to 293~K and $4~10^{-4}$~s$^{-1}$.

\begin{figure}
\setlength{\unitlength}{1cm}
\begin{center}
\includegraphics[width = 12 cm]{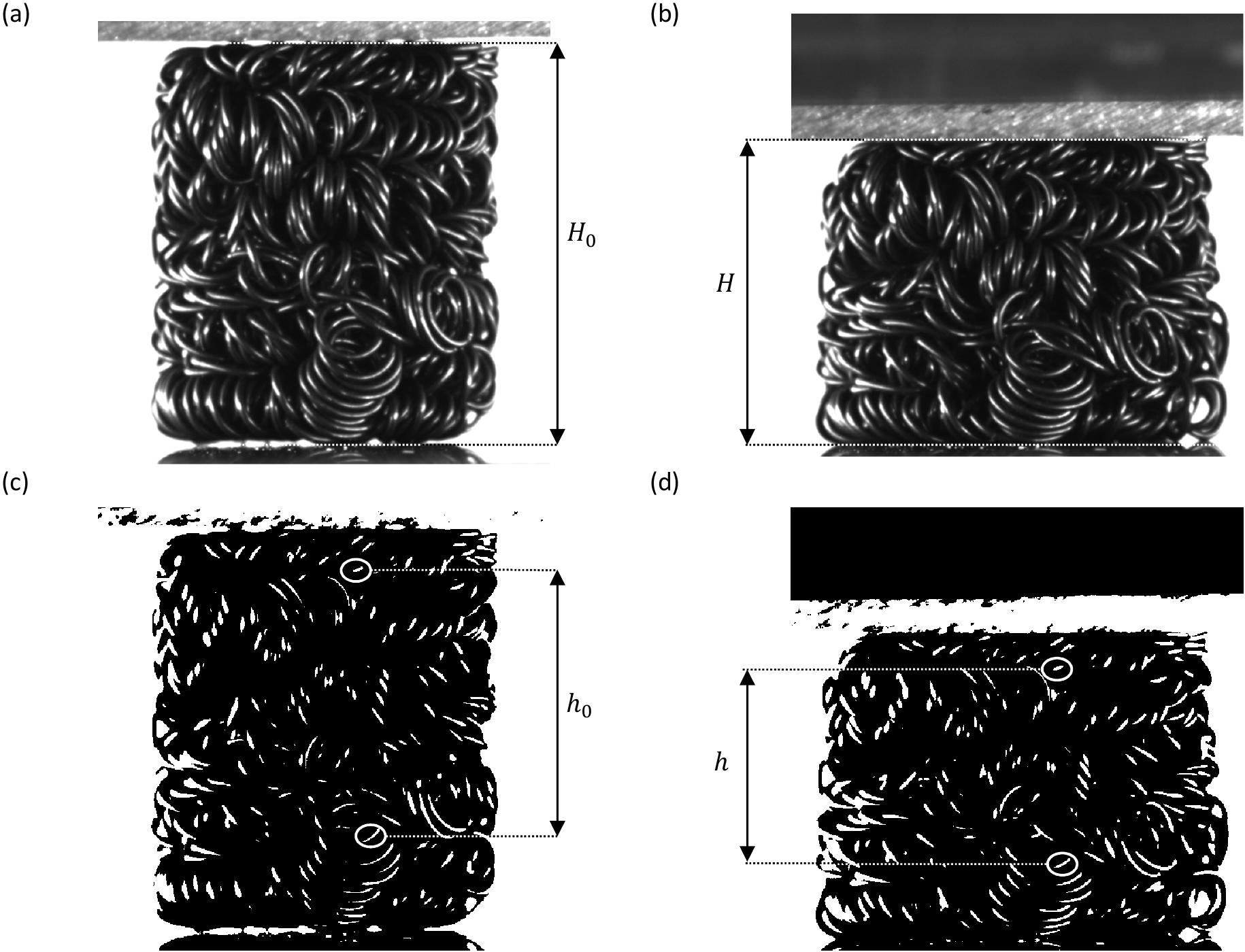} 
\caption{\label{fig_tracking} Illustration of the particle tracking method used to assess the local compressive strain: grey scale (a,b) and segmented (c,d) pictures of the sample before (a,c) and during (b,d) compression.}
\end{center}
\end{figure}

To avoid errors on the strain measurements induced by edge effects such as friction between the sample and the plates or inhomogeneities on the outside of the samples, a local strain measurement was developed using images of the sample taken during the tests with a CCD camera (Jai Pulnix RM-4200GE, spatial resolution 2048$\times$2048). As illustrated in Fig. \ref{fig_tracking}, a segmentation of the grey scale images (Fig. \ref{fig_tracking}(a,b)) together with a particle tracking procedure (developed with ImageJ and Matlab) were used to obtain the coordinates of the centers of mass of the enlightened parts of the fiber (Fig. \ref{fig_tracking}(c,d)). By following the image sequences, we defined a gauge zone in the center of the samples, within which the strain could be assessed. The enlightened parts of the fibers that were as close as possible from the borders of this zone (in the initial configuration) were automatically detected and followed during the compression cycles to determine the initial ($h_0$) and  current ($h$) heights of the gauge region, from which we determined the local Hencky axial strain $\varepsilon=\ln (h/h_0)$.
%

\section{Wire properties after processing}
\label{par:wire}

\begin{figure}
\setlength{\unitlength}{1cm}
\begin{center}
\includegraphics[width = 12 cm]{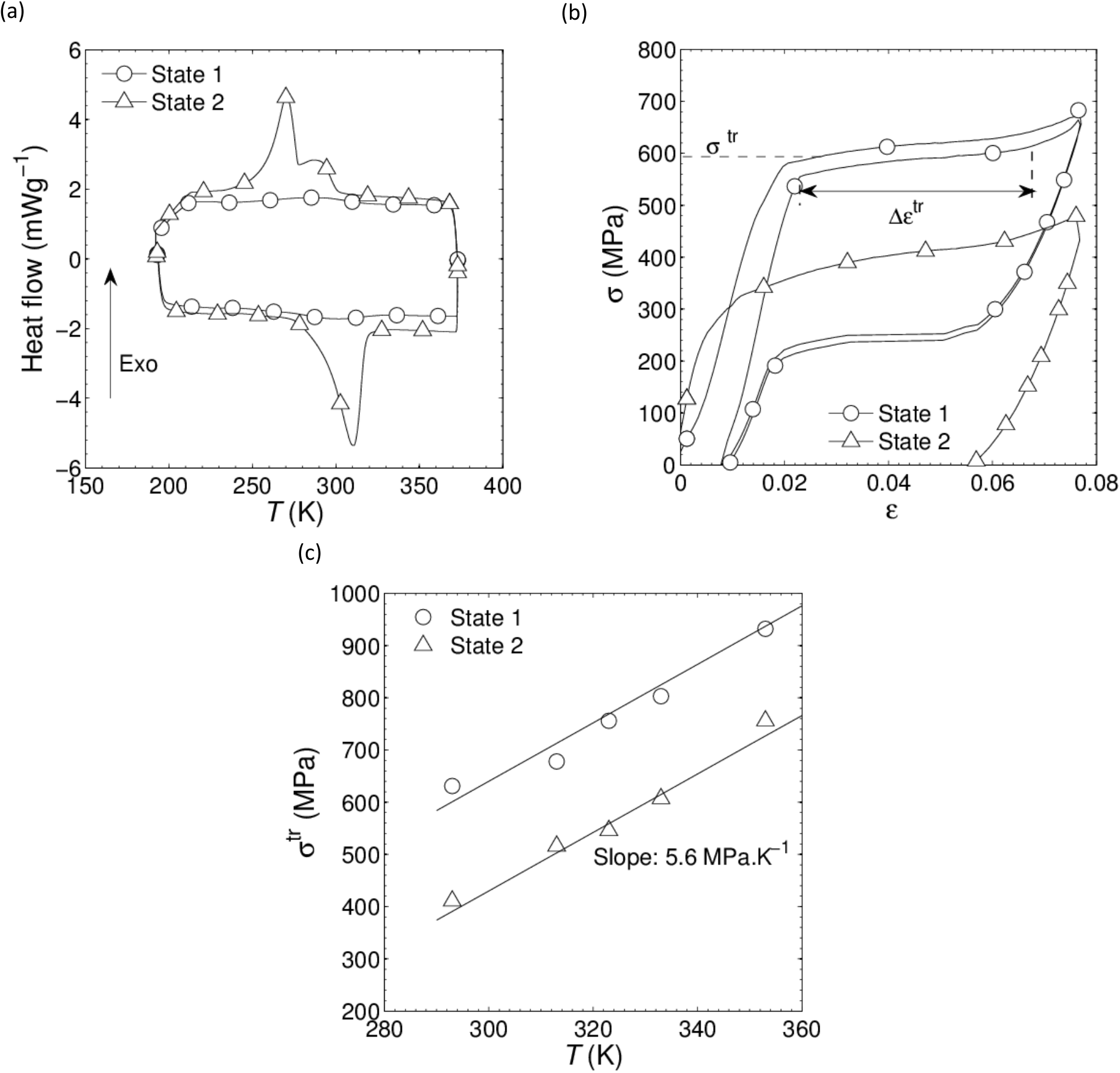} 
\caption{\label{fig_prop_wire} DSC curves of small portions of the entanglements in \textit{states 1} and \textit{2} (a), tensile stress-strain curves of two straight wires subjected to same thermal history as the entanglements (b), and evolution of the forward stress-induced transformation stresses $\sigma^{tr}$ with temperature (c).}
\end{center}
\end{figure}

As shown in Fig. \ref{fig_prop_wire}(a), within the investigated temperature range, DSC measurements for the entangled wires in \textit{state 1} exhibit a single transformation peak upon cooling and heating, presumably  associated with forward A-R (Austenite to R-Phase) and reverse R-A martensitic transformations These peaks are very flat, with latent heats $\Delta H^{A-R}\approx \Delta H^{R-A} \approx 2$~J~g$^{-1}$. Conversely, in \textit{state 2}, the wires exhibit two stage A-R-M (Austenite to R-phase to Martensite) and M-R-A transformations with pronounced transformation peaks and latent heats ($\Delta H^{A-R-M}\approx \Delta H^{M-R-A} \approx 16$~J~g$^{-1}$). Upon heating, the transformation end temperature, $A_f$, is close to 328~K.

As expected from the DSC measurements, the room temperature wire tensile behavior in \textit{state 1} is superelastic (see Fig. \ref{fig_prop_wire}(b)): upon loading, after an initial apparent elastic regime, the stress-strain curve shows a high R-M transformation stress, $\sigma^{tr}$ $\approx$ 550 MPa, and a transformation plateau within a strain range $\Delta \varepsilon^{tr}\approx 0.05$. Unloading leads to the reversed process, with a pronounced stress hysteresis and a small residual strain. Conversely, wires in\textit{ state 2} are ferroelastic, with a large residual strain upon unloading due to the unrecoverable strain produced during loading by the stress-induced A-R-M and A-M forward transformations. Note however that the transformation strain may be recovered  by stress-free heating above $A_f$. Also, as seen in Fig. \ref{fig_prop_wire}(c), the transformation stresses recorded in both states at the onset of the stress-induced forward transformation, $\sigma^{tr}$, follows a Clausius-Clapeyron like relation, with a slope close to 5.6~MPa.K$^{-1}$, in agreement with values commonly reported for similar heat treated NiTi alloys \citep{Stachowiak88,Pelton00,Favier06}.

\section{Mesostructure of the entanglements}
\label{par:micro}
%

Fig. \ref{fig_tomo} summarizes the information gathered on the entanglement mesostructure. We focus here on 2 representative samples with mean relative densities $\bar\phi=$0.3 and 0.36. Fig. \ref{fig_tomo}(a) shows a cut through a tomographic reconstruction, where we see a complex architecture that may be described at two levels: at short range, the structure is made of a quasi-ordered spring, but at longer range, the spring is entangled with itself in a disordered fashion. As illustrated from the two profiles of the normalized relative density in Fig. \ref{fig_tomo}(c-d) ($\phi^*=\phi/\bar\phi$, $\phi$ and $\bar\phi$ being the local and mean relative densities), the fiber content inside the samples is close to homogeneous, except slight decreases of $\phi^*$ close to the surface of samples. This is in contrast with systems obtained after entangling a straight wire where the density strongly increases on the outside of the samples \cite{Courtois12}: in this situation, the reason is that the fiber decreases its curvature by moving to the outside. Here, this relaxation is not possible because the high tortuosity of the spring leads to the formation of numerous contacts in the spring ball (step (ii) in Fig. \ref{fig_processing}) that efficiently lock the fiber.

The pore size distribution shown in Fig. \ref{fig_tomo}(f) exhibits a broad peak at $\approx 1300$~$\mu$m for $\bar\phi=0.3$. As expected, increasing the fiber content to $\bar\phi=0.36$ creates smaller pores,  with a sharper peak shifted down to $\approx 750$~$\mu$m. The peak pore size may thus be tuned by simply changing the length of the initial wire (and thus the fiber content $\bar\phi$) without changing any of the other processing parameters. This will also affect the mechanics of the entanglements (see next Section). It is also interesting to notice that the pore size distribution could also be shifted down at fixed fiber content (and presumably fixed mechanical properties) simply by reducing both the wire diameter and the diameter of
the threaded bar used to coil the springs. Thus, combined with the open-pore nature of the entanglement, the resulting mesostructure would fit values commonly admitted for bone implants, \textit{i.e.} mean pore size between 100 and 600~$\mu$m \citep{Bansiddhi08}. 

Distributions of the mean curvature are reported in Fig. \ref{fig_tomo}(g). They display narrow peaks near 0.55~mm$^{-1}$ and 0.52~mm$^{-1}$ for $\bar\phi=$0.3 and 0.36, respectively. These values are close to the principal curvature $\bar\kappa_0\approx$~0.57~mm$^{-1}$ of the non-entangled spring. Thus, the moderate stretching and folding of the spring induced during steps (ii) and (iii) of the manufacturing process do not induce noticeable changes of this mesostructure parameter.   

The diagonal values of the orientation tensors $\matr{A}$ and $\matr{B}$ are nearly identical and close to 1/3 (see the data given in Fig. \ref{fig_tomo}(e) for the sample with $\bar\phi=$0.36), confirming that on average, the entanglement is isotropic, without preferred orientation for the wire centerline or its self-contacts. 

The estimated number of full contacts per unit fiber length, $n$, is close to 0.40~mm$^{-1}$ and 0.50~mm$^{-1}$ for $\bar\phi=$0.27 and 0.36, respectively. In other words, the two scanned samples exhibit respectively 2608 and 3874 full contacts. This emphasizes both the very high connectivity of the entanglements and its increase with fiber content $\bar\phi$. Besides, the measured ratio $r=n(\bar\phi=0.3)/n(\bar\phi=0.36)$ is close to 0.83. It is worth mentioning that this ratio is consistent with the ratio $r$ predicted by the excluded volume theory for isotropic networks of straight fibers with circular cross section \citep{Doi78,Ranganathan91,Toll93,Orgeas12}:
\begin{equation}
r(\phi_1,\phi_2)={\phi_1}/{\phi_2},
\end{equation} 
which yields $r=0.81$ for $\phi_1=$0.3 and $\phi_2=$0.36, again emphasizing the effect of the fiber content on the entanglement connectivity.

Lastly, as shown in Fig. \ref{fig_tomo}(h), the distributions of contact lengths are wide with a predominant contact population marked by a distinct peak, but also a distribution of short contacts, down to zero length. 
It is also worth noting that the predominant contact length respectively peaked close to $400~\mu$m and $700~\mu$m for $\bar\phi=$0.3 and 0.36: increasing the fiber content yields to an increase of both the number of full contacts (see previous point) and their length.
%

\section{Thermomechanical behavior}\label{par:resu}
\subsection{Typical results for samples in \textit{state 1}}
\begin{figure}
\setlength{\unitlength}{1cm}
\begin{center}
\includegraphics[width = 6 cm]{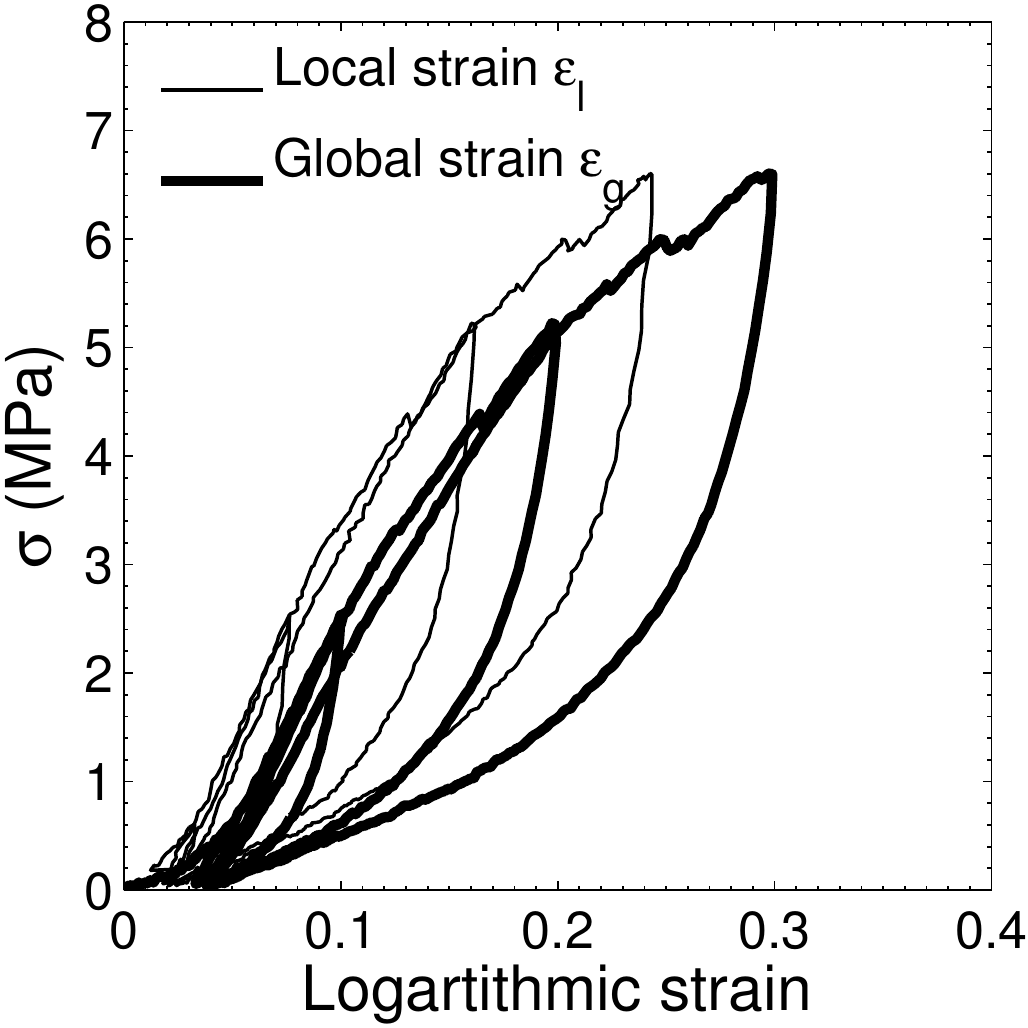} 
\caption{\label{fig_typical} Typical stress-strain curves obtained during compression load-unload cycles for a sample a sample in \textit{state 1} ($\bar\phi = 0.36$, reference testing temperature and strain rate).}
\end{center}
\end{figure}

Fig. \ref{fig_typical} shows typical stress-strain curves obtained during compression load-unload cycles at the reference testing temperature and strain rate. Here the stress is shown as a function of both the global and local strains. At the sample scale, Fig. \ref{fig_tracking} shows that the deformation proceeds without pronounced friction effects at the compression plates since the initial cylindrical shape is preserved even at finite deformation. However, we have seen in Fig. \ref{fig_tomo} that the relative density is not fully homogeneous, in particular near the top and bottom of the samples. We may thus expect more pronounced deformations in this regions. Indeed, we find in Fig. \ref{fig_typical} that the global strain measured at the level of the structure is noticeably larger than the local strain, which is measured in a gauge region near the center of the sample. We ascribe this difference to larger deformations near the top and bottom of the samples. For this reason, we will consider only the local strain in the following.

As shown in Fig. \ref{fig_typical}, the loading curve is practically recovered during successive cycles. The fibrous entanglement thus exhibits no stress accommodation, which is rarely observed in non-bonded fibrous materials, where irreversible rearrangements of the fiber segments, or, in case of elastoplastic (resp. brittle) fibers, plastic deformation (resp. rupture) of fibers, occur during cycling \citep{Robitaille98,Poquillon05,Zhang13}. This marked stability of the entangled structure arises from the numerous self-contacts and the fact that during deformation, sufficient internal back stresses arises in the wire to reverse during unloading the rearrangements induced upon loading, at least at the macroscopic level of the structure.

The stress-strain loading curve exhibits a marked strain hardening, even at elevated compression strains, which is interesting for structural applications. This was observed independently of the state, fiber content, temperature and strain rate. This phenomenological feature is different from foams \citep{Gibson01}, which exhibit stress plateaux. The reason is the continuous formation of new contacts during compression, which stiffen the structure. The stress-strain curves in Fig. \ref{fig_typical} start with a low slope, at least during the first cycle, which is typical of non-bonded fibrous systems where the fibers rearrange to adapt to the applied strain. Then, the strain-stiffening rate increases rapidly up to about $\varepsilon = 0.1$. This behavior is related to a consolidation of the structure \citep{Toll98,Poquillon05,Masse06,Picu11}, the principal underlying mechanisms being a rearrangement of the fibrous mesostructure, an increase of the number and/or  surface of fiber-fiber contacts, which in turn restrains the motion, rearrangement and deformation of fiber segments \citep{VanWyk46,Toll98}. For strains above about 0.1, the stress increases less rapidly, which is unusual for compressed fibrous materials made of elastic fibers. In the case of elastoplastic (or brittle) fibers, this phenomenon is related to the plasticity (resp. rupture) of fibers during loading \citep{Toll98,Tan09,Courtois12}. For NiTi wires, it should rather be ascribed to the forward martensitic transformation induced in the vicinity of contact zones where the wire is significantly bent.

Fig. \ref{fig_typical} also shows upon unloading a nearly full recovery of the imposed strain, even when the maximum local strain reaches 0.3 (corresponding to a global strain $\approx 0.4$). The strain recovery is accompanied with a noticeable hysteresis. To the best of our knowledge, such macroscale superelastic behavior has never been observed at elevated strains in porous/fibrous metals, even with similar mesostructures. Indeed, \citet{Tan09} and \citet{Zhang13} observed a nearly full compression strain recovery for entangled elastoplastic Al alloy monofilaments but with a maximum applied global strain of 0.13 only to avoid wire plasticity. The present macroscale superelastic behavior is induced by the elastic energy stored in the wire upon loading and released upon unloading. It is also probably related to stress-induced reversible martensitic transformations taking place during the unloading phase. Both phenomena offer sufficient driving force to the wire to reverse the deformation mechanisms that occurred upon loading, despite significant mesostructure rearrangements and friction forces at contacts. 
%

%
\subsection{Influence of the strain rate and testing temperature}

\begin{figure}
\setlength{\unitlength}{1cm}
\begin{center}
\includegraphics[width = 12 cm]{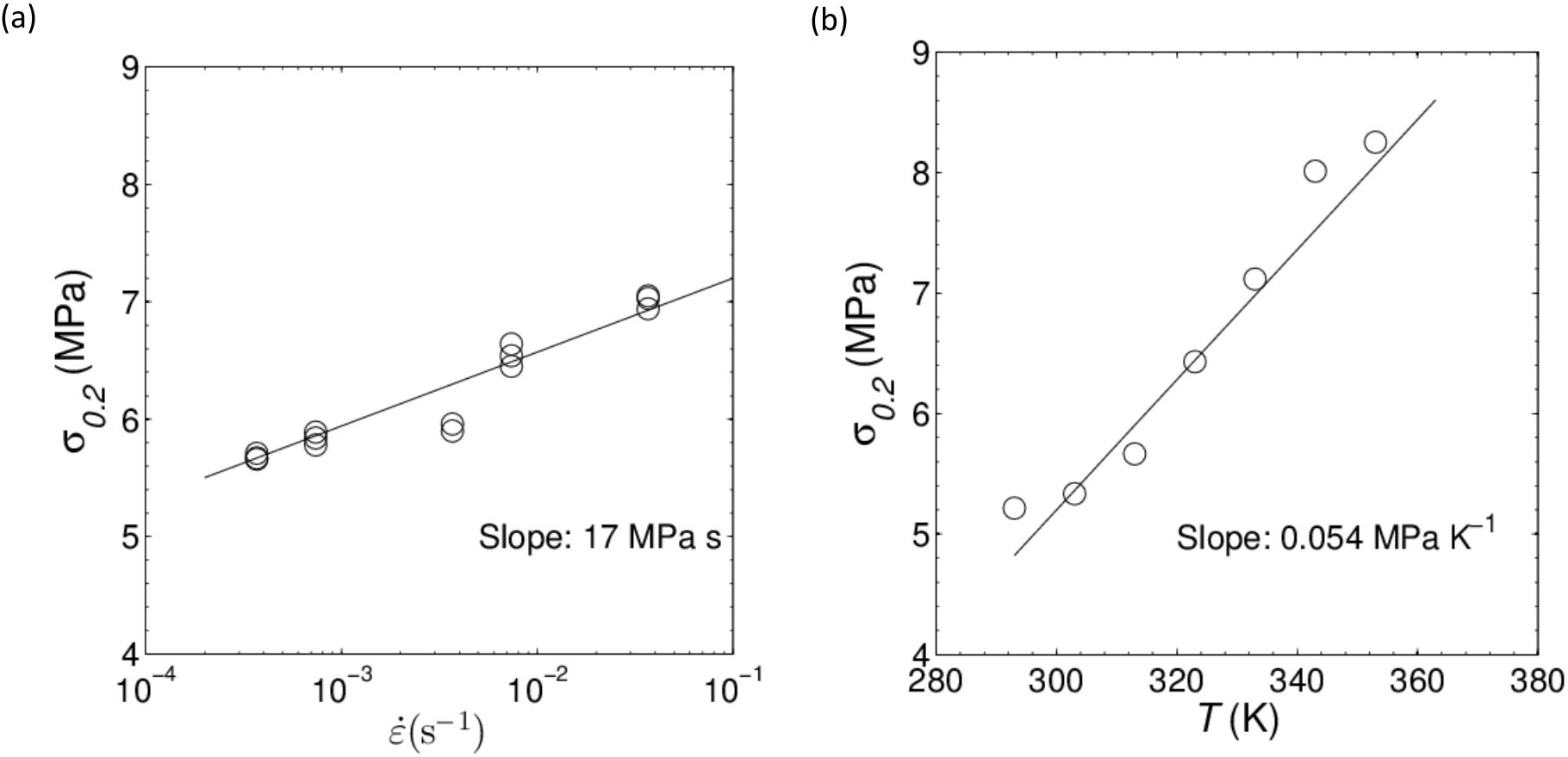} 
\caption{\label{fig_inf_T} Evolution of the compression stress $\sigma_{0.2}$ with strain rate $\dot\varepsilon$ ($\bar\phi$=0.3 and $T$=303K) (a) and testing temperature $T$ ($\dot\varepsilon$=4 10$^{-4}$ s$^{-1}$) (b).}
\end{center}
\end{figure}

Fig. \ref{fig_inf_T} shows the evolution of $\sigma_{0.2}$, the stress at $\varepsilon=$0.2, as a function of the strain rate (Fig. \ref{fig_inf_T}(a)) and testing temperature (Fig. \ref{fig_inf_T}(b)). The strain rate has a weak influence, with an increase of $\sigma_{0.2}$ by less than 20\% when the strain rate is increased by 3 orders of magnitude. This weak increase may be related to the occurrence of thermomechanical couplings during stress-induced transformation in the wire \citep{Shaw97,Favier06,Sun10}. 
%
Fig. \ref{fig_inf_T}(b) shows the evolution of $\sigma_{0.2}$ for a sample compressed at $\dot\varepsilon~=~4~10^{-4}$~s$^{-1}$ at temperatures from 293 to 353~K. Obviously, the effect of  the temperature is also limited, with a slope of about 0.05 MPa.K$^{-1}$, more than two orders of magnitude smaller that for a single wire (see Fig. \ref{fig_prop_wire}). Thus, results emphasized in Fig. \ref{fig_inf_T}(a) and (b) lead us to conclude that during the investigated compression tests, a very small volume fraction of the wire is transformed. 

\subsection{Influence of the fiber content and the maximum compression strain}
\begin{figure}
\setlength{\unitlength}{1cm}
\begin{center}
\includegraphics[width = 12 cm]{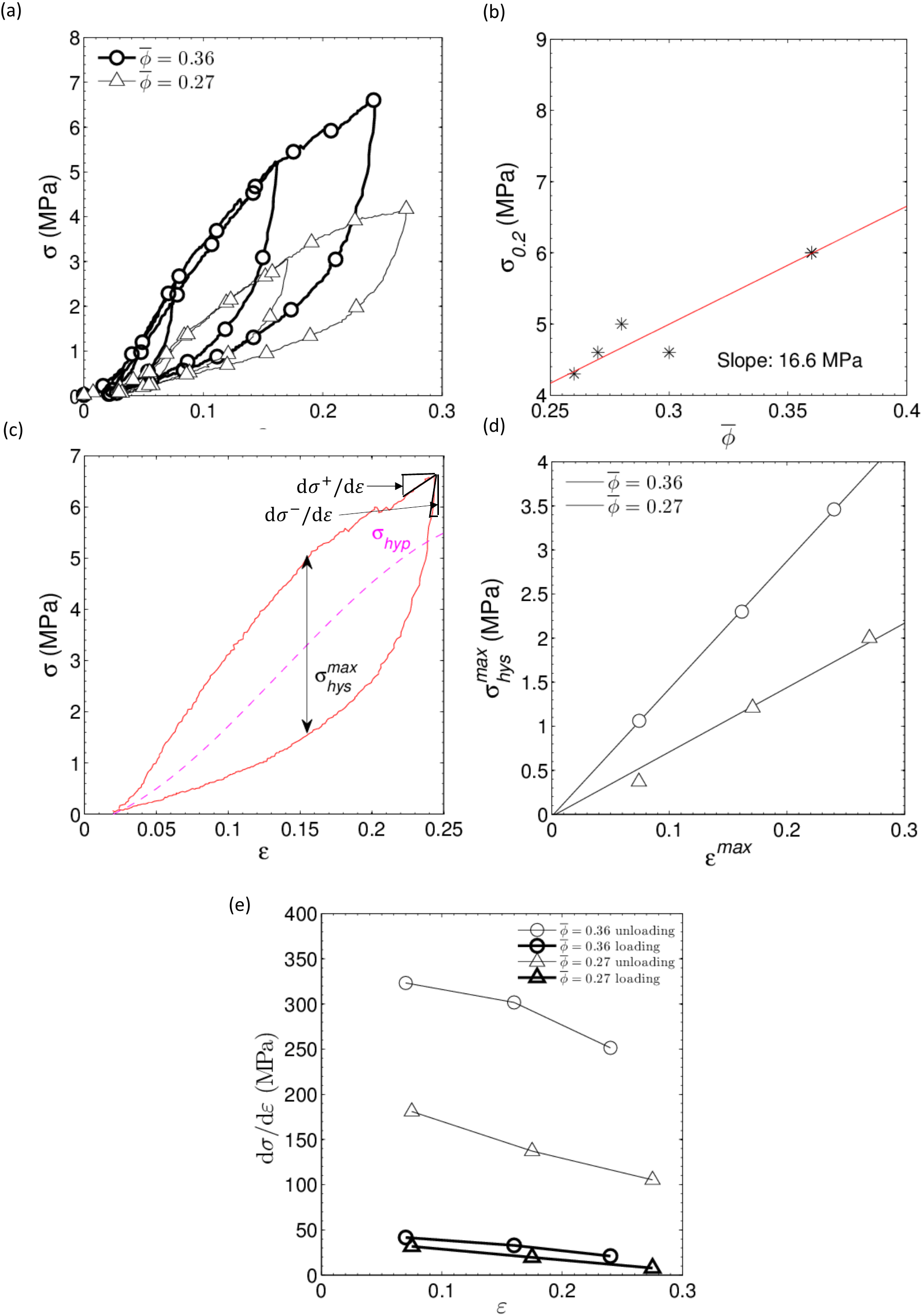} 
\caption{\label{fig_inf_phi} Influence of the fiber content $\bar\phi$ and maximum compression strain $\varepsilon^{max}$ on stress-stress curves (a), stress $\sigma_{0.2}$ (b), maximum stress hysteresis $\sigma_{hys}^{max}$ (c-d) and tangent moduli $\mathrm{d}\sigma^+/\mathrm{d}\varepsilon$ and $\mathrm{d}\sigma^-/\mathrm{d}\varepsilon$ (e).}
\end{center}
\end{figure}

Fig. \ref{fig_inf_phi} shows the influence of the initial fiber content and of the maximum compression strain imposed in a given load-unload cycle, on the mechanical behavior of the entanglements. As illustrated in Fig. \ref{fig_inf_phi}(a), the higher the fiber content, the higher the stress level upon loading, the larger the stress hysteresis and the larger the tangent moduli of stress-strain curves. These trends are often observed in non-bonded fibrous materials \citep{Toll98,Poquillon05,Masse06,Courtois12}.

The evolution of stress levels with fiber content is more precisely emphasized in Fig \ref{fig_inf_phi}(b). For the present range of fiber contents, the increase of $\sigma_{0.2}$ is limited: from $\bar\phi=$~0.27 to 0.36, $\sigma_{0.2}$ raises only from 3.6 to 6~MPa. This trend is directly related to the increase of the number and length of self-contacts, as emphasized in the last Section (Fig. \ref{fig_tomo}): the larger the number and length of self-contacts, the more difficult the rearrangements and the higher the macroscale stress \citep{VanWyk46,Toll98,Masse06,Latil11,Picu11}.

The maximum hysteresis stress $\sigma^{max}_{hys}$ of a given load-unload cycle was estimated by measuring the maximum stress gap between loading and unloading, as illustrated in Fig. \ref{fig_inf_phi}(c). As shown in Fig. \ref{fig_inf_phi}(d), $\sigma^{max}_{hys}$ is approximately twice larger when $\bar\phi=$0.36 than when $\bar\phi=$0.27. Besides, a nearly linear increase of $\sigma^{max}_{hys}$ with $\varepsilon^{max}$ is also noticed. These two trends can be explained by three underlying phenomena. First, the hysteresis associated with forward and reverse stress-induced martensitic transformations increases with the maximum strain applied during a cycle \citep{Cory85,Orgeas97,Orgeas04}. However, this phenomenon should be of second order given the small volume fractions of transformed wire estimated above. The second possible mechanism is the rearrangement of the fibrous mesostructure: although the entanglements exhibit a superelastic behavior without marked residual strain at the macroscopic level, the rearrangements at the wire level follow different paths during loading and unloading (see next Subsection). Lastly, there is the friction induced by forces arising at self-contacts, which is closely linked with the previous phenomenon. 

Tangent moduli just before, $\mathrm{d}\sigma^+/\mathrm{d}\varepsilon$, and just after, $\mathrm{d}\sigma^-/\mathrm{d}\varepsilon$, the return point of a given load-unload cycle (see Fig. \ref{fig_typical}(b)) were estimated ($\mathrm{d}\sigma^-/\mathrm{d}\varepsilon$ estimates the Young modulus $E$ of the entanglements, as often proposed for foams or fibrous materials). Their evolution with $\varepsilon^{max}$ are reported in Fig. \ref{fig_typical}(e). 
As expected, $\mathrm{d}\sigma^+/\mathrm{d}\varepsilon$ is much lower than $\mathrm{d}\sigma^-/\mathrm{d}\varepsilon$, showing that, as expected, non-elastic mechanisms, such as wire rearrangements or friction as mentioned above, are involved during loading. Moreover, the measured moduli increase with $\varepsilon^{max}$, which could be correlated to a possible increase of the actual fiber content during loading and/or the increase of the number/surface of fiber contacts (induced by the possible increase of the actual fiber content or by the rearrangement of the entanglements). 
Furthermore, a noticeable increase of $E$ with the initial fiber content $\bar\phi$ is noticed: this is directly connected to the increase of the number of fiber contacts with $\bar\phi$ (see section \ref{par:micro}, \citep{Toll98}).  

Lastly, the Young moduli estimated from $\mathrm{d}\sigma^-/\mathrm{d}\varepsilon$ in Fig \ref{fig_inf_phi}(b) strongly depend on the fiber content and less on the applied strain. We find Young moduli broadly between 100 and 350 MPa. These are lower than often reported for metal foams with similar relative densities (in general above 1 GPa \citep{Gibson01}). Nonetheless, the Young moduli found here are within the range of 10-3000 MPa estimated for trabecular bones \citep{Kopperdahl02,Gibson01,Geetha09}, and very close to values given for porous NiTi materials used for biomedical implants \citep{Prymak05}. The present architecture may therefore be a potential candidate for biomedical applications, with the very interesting advantage of being superelastic and highly adaptable to applied loads.

\subsection{Hysteresis and Discrete Memory}

\begin{figure}
\setlength{\unitlength}{1cm}
\begin{center}
\includegraphics[width = 12 cm]{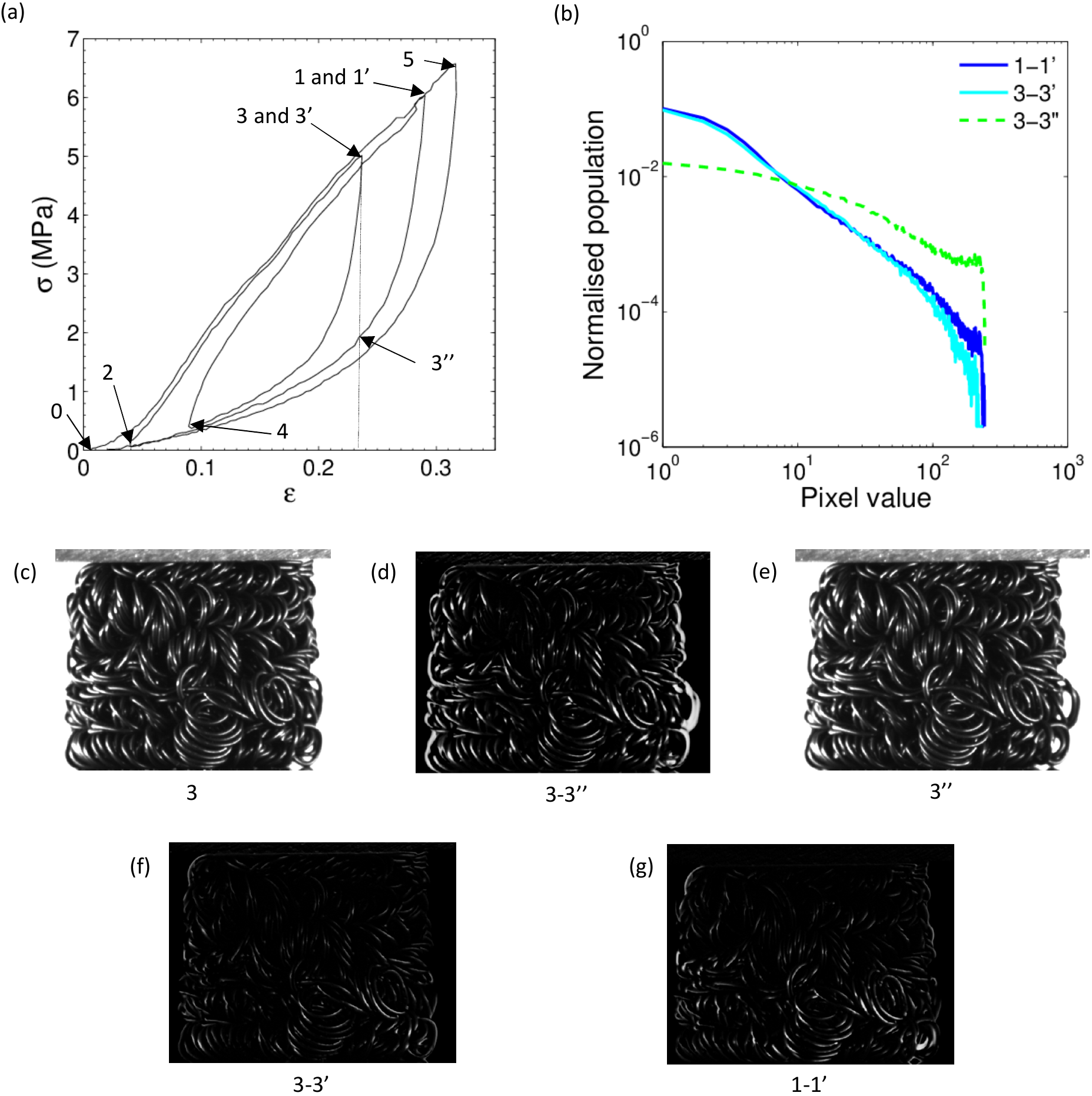} 
\caption{\label{fig_dicrete} Complex cycling 0-1-2-3-4-5 performed on an entanglement to illustrate its discrete memory both on stress-strain diagram (a) and on the mesostructure (b-g): images taken at points 1, 1', 3 (c), 3' and 3'' (e) were used to compute the image differences 3-3'' (d), 3-3' (f) and 1-1' (g), the grey scale histograms of which are plotted in (b).}
\end{center}
\end{figure}

The hysteretic behavior of the entanglements is further explored in Fig. \ref{fig_dicrete}. This stress-strain diagram was obtained using a specifically designed cycling. A first major load-unload loop noted 0-5-0 in Fig. \ref{fig_dicrete} was performed from $\varepsilon=0$ to $\varepsilon^{max}=0.3$ then back to $\varepsilon=0$.  Then, a partial loop cycling was started at the return point 1 and continued following the numerical order of the return points marked in Fig. \ref{fig_dicrete}, \textit{i.e.} 1-2-3-4-5. We have also reported in Fig. \ref{fig_dicrete} images extracted from the video recorded during the experiment. In Fig. \ref{fig_dicrete}(d), we computed the difference 3-3'' from the image of the sample at the return points 3 and 3'' (Fig. \ref{fig_dicrete}(c)), \textit{i.e.} two points with the same strain $\varepsilon$ but a different mechanical history. In this image, the larger the difference at a given pixel, the brighter the pixel. The same procedure was carried out between points 3 and 3' (Fig. \ref{fig_dicrete}(f)) and 1 and 1' (Fig. \ref{fig_dicrete}(g)). The grey scale histograms deduced from these differences are shown in Fig. \ref{fig_dicrete}(b). 

As clearly evidenced from Fig. \ref{fig_dicrete}(a), the stress-strain curve of the section 4-5 passed through all the previous return points, thus closing all the incomplete partial loops, 3-4-3 and 1-2-1 initiated during the pathway 1-2-3-4. Fig. \ref{fig_dicrete}(d) also reveals some mesotructure differences between points 3 and 3'', although these two configurations have the same compression local strain. Such a difference is also emphasized with the rather flat grey scale histogram of Fig. \ref{fig_dicrete}(b). Both figures highlight the rearrangement of the fiber segments between the loading and unloading phases of a cycle. By contrast, images 3 and 3', as well as 1 and 1' nearly coincide, despite the complex pathways used to go from 3 to 3' and from 1 to 1'.  

Similar behavior was observed in superelastic or ferroelastic bulk NiTi or CuZnAl alloys \citep{Orgeas97,Orgeas04}. As for these systems, the present entanglements  follow the concept of ``erasable micromemory'' also known as the ``return point memory'' or ``discrete memory'' effect \citep{Guelin80,Vergut87,Ortin92}. The re-assembling of the return points along the path 4-5 suggests that the mechanical behavior of the entanglements is ruled by a part of its previous deformation history. More precisely, once a subloop is initiated, the entanglement "remembers" the subloop starting/return point, namely its reference stress-strain state and it also remembers its reference mesostructure until the subloop is closed. Moreover, once the subloop is closed, it is ``forgotten'' in the way that it does not affect the parent loop within which it was performed. This is illustrated along the section 4-5 after closing subloop 3-4-3 at return point 3: the stress-strain curve follows its parent loop 1-2-1 as if no subloop had been carried out before. Similarly, the stress-strain curve follows the major loop 0-5-0 as if subloop 1-2-1 and 3-4-3 had been erased from the entanglement mechanical history. This behavior could be potentially interesting for practical applications. From a fundamental standpoint, it raises questions about the mechanics of such non-bonded entanglements that are able to memorize some of their mesostructure configurations, even after complex trajectories and rearrangements.   
\subsection{Hysteresis and Energy Dissipation}
\begin{figure}
\setlength{\unitlength}{1cm}
\begin{center}
\includegraphics[width = 12 cm]{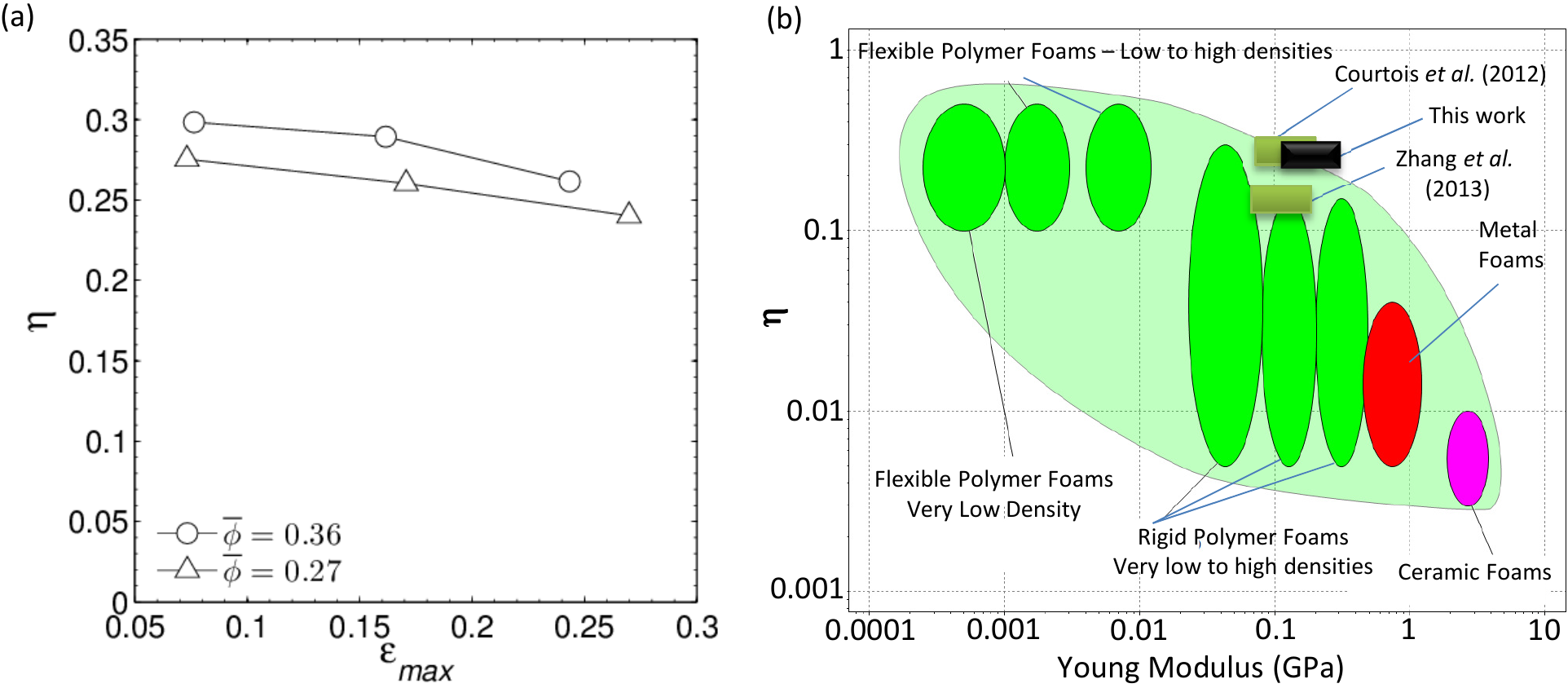} 
\caption{\label{fig_amortis} Evolution of the loss factor $\eta$ with strain magnitude $\varepsilon^{max}$ for two fiber contents $\bar\phi$ (a). Material map of the loss factor $\eta$ versus the Young modulus (from CES \citep{ces2013})) (b): metallic entanglements are represented by rectangles, with the present NiTi architectures in black.}
\end{center}
\end{figure}

The observed hysteresis is related to dissipative mechanisms. To estimate these effects, the energy loss $E_{hys}$ induced during (sub)loops was computed and compared with the macroscale elastic energy $E_{el}$  stored and released during such sequences. For each closed (sub)loop, $E_{hys}$ was estimated from the following expression:
\begin{equation}
E_{hys}=\oint \sigma\mathrm{d}\varepsilon.
\end{equation}
The elastic energy  $E_{el}$ was calculated by assuming that the overall stress produced by the entanglement can be seen as the sum of two contributions: a hysteretic stress $\sigma_{hys}$ and a hyperelastic stress $\sigma_{hyp}$, defined as the middle (neutral) line of the major hysteresis loop (see Fig. \ref{fig_inf_phi}(b) for an illustration) \citep{Guelin80,Favier10,Zhang13}. Hence, for a (sub)loop carried out between two strain states $\varepsilon_1$ and $\varepsilon_2$, $E_{el}$ is calculated as:
\begin{equation}
E_{el}=\int_{\varepsilon_1}^{\varepsilon_2} \sigma_{hyp}\mathrm{d}\varepsilon.
\end{equation}
The loss factor $\eta$, \textit{i.e.} the energy dissipation coefficient, is then computed as \citep{Zhang13}:
\begin{equation}
\eta=\frac{E_{hys}}{\pi E_{el}}.
\end{equation}
Fig. \ref{fig_amortis}(a) reports the evolution of $\eta$ with the strain magnitude $\Delta\varepsilon=\varepsilon_2-\varepsilon_1$  deduced from stress-strain diagrams such as shown in Fig. \ref{fig_typical} ($\varepsilon_1=0$ and $\varepsilon_2=\varepsilon^{max}$). We can see that $\eta$ depends on $\bar\phi$ than on $\varepsilon^{max}$. As illustrated by the material map plotted in Fig. \ref{fig_amortis}(b), the values of $\eta$  found here between 0.25 and 0.3 are higher than most of metallic foams, because of the increased  friction at contacts as well as the forward and reverse stress-induced martensitic transformations. Besides, with a much higher Young modulus, the entanglements have loss factors as high as the best polymeric flexible foams. Finally,  the processed materials display loss factors similar to other entangled mesostructures \citep{Courtois12,Zhang13} whereas they can be deformed reversibly at higher compression strains. Such architected structures are thus potentially relevant design solutions for load bearing and damping components. 
\subsection{Samples in state 2: ferroelasticity and shape memory effect}

\begin{figure}
\setlength{\unitlength}{1cm}
\begin{center}
\includegraphics[width = 12 cm]{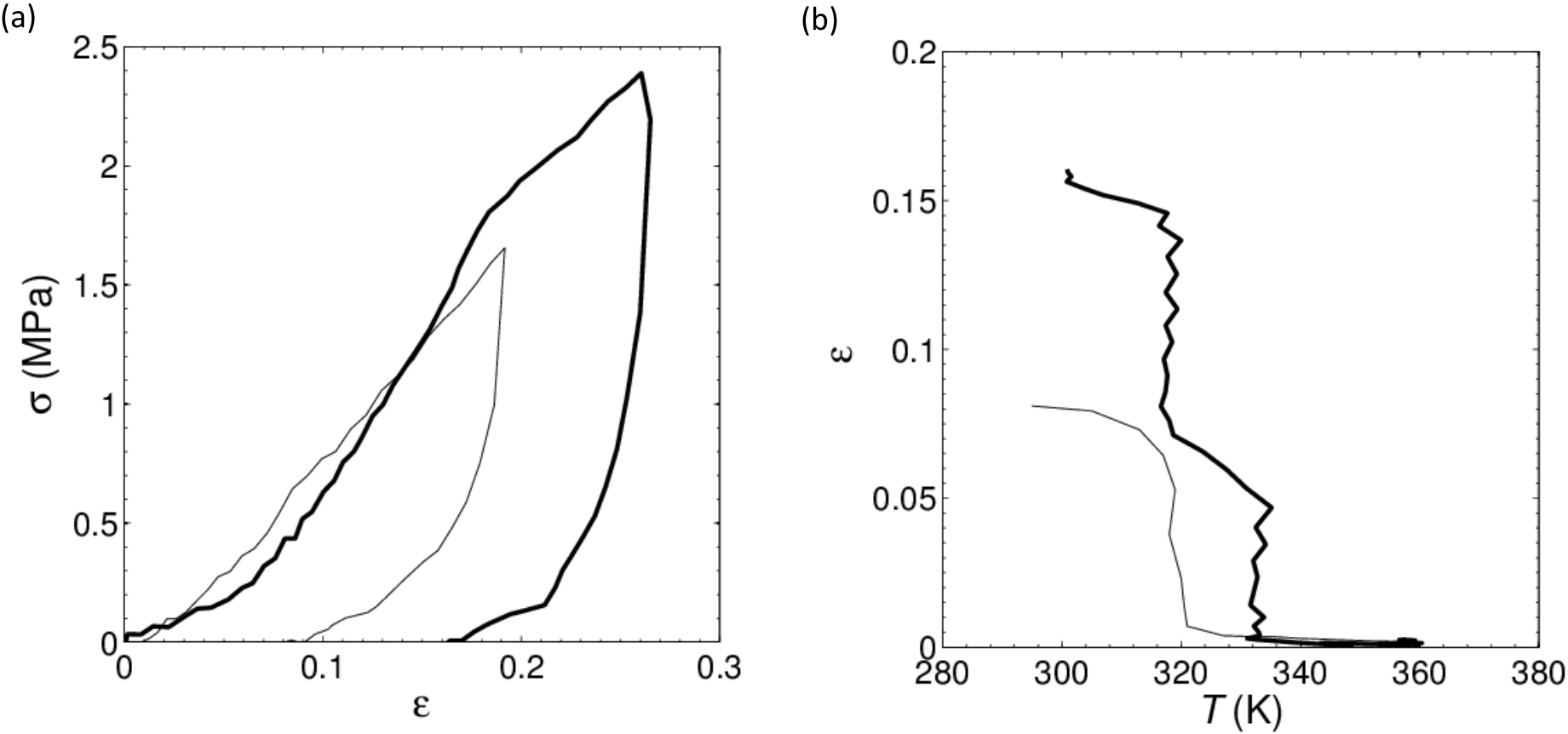} 
\caption{\label{fig_ferro} Stress-strain curves (a) and strain-temperature curves (b) showing the ferroelastic behavior of the entanglements in \textit{state 2} and its associated shape memory effect observed upon stress-free heating after the load-unload sequence shown in (a) ($\bar\phi$=0.3).}
\end{center}
\end{figure}

By simply changing the heat treatments during sample processing and going from \textit{state 1} to \textit{state 2},  it is possible to tune the thermomechanical behavior of the entanglements, as illustrated in Fig. \ref{fig_ferro}. The two stress-strain curves in Fig. \ref{fig_ferro}(a) were obtained with samples in \textit{state 2} compressed at two different maximum strains at the reference temperature and strain rate ($\bar\phi=0.3$). The mechanical response is drastically modified compared with the stress-strain curves in \textit{state 1} (see Fig. \ref{fig_inf_phi}(a)). First, the stress levels are much lower. Second, the superelastic behavior of \textit{state 1} is replaced by a ferroelastic behavior with large residual strains. The residual strain reaches 0.08 and 0.16 for maximum imposed strains of 0.2 and 0.27, respectively. From a phenomenological standpoint, the shape of stress-strain curves is close to those observed for similar entanglements with elastoplastic wires \citep{Zhang13}. However, for the present NiTi wire, the shape of the stress-strain curve is not controlled by the wire ductility but rather by the low value of the transformation stress required to induce martensite within the wires (see Fig. \ref{fig_prop_wire} (b)). Thus, in \textit{state 2}, the energy stored in the wire upon loading is not high enough to overcome the configurational and friction forces required to push back the rearrangements and the wire deformation is highly irreversible upon unloading.

The large residual strains induced at room temperature can be fully recovered simply by a stress-free heating up to 353~K, as evidenced in Fig. \ref{fig_ferro}(b): the samples come back to their initial shapes upon heating, thus exhibiting a large, full and thus unprecedented shape memory effect. Upon heating, the energy stored in the transformed zones of the wire is progressively increased according to the Clausius-Clapeyron relation and become sufficient to reverse the transformation in these zones, pushing back the rearrangements induced during the deformation at room temperature.  
%

%
\section{Concluding Remarks}
\label{par:conc} 

By entangling a single NiTi spring followed by a compaction in a close die, we were able to manufacture homogeneous porous NiTi materials with controlled mesostructures and physical properties. The manufacturing method is versatile enough to tune easily the process-induced mesostructures. For examples, by changing the compaction rate or the initial wire length, samples with different fiber contents may be obtained. Also, a simple scaling of the spring dimensions would allow to tailor the pore size distribution and changing the shape of the pre-compacted entanglement may also lead to orientated specimen with anisotropic properties. Also, compared with other methods used to produce NiTi porous materials, another advantage of the present processing route is the use of only low temperature heat treatments. This avoids the formation of an oxide skin, which is detrimental to the mechanical properties of NiTi. Another advantage of the present method is that the temperature and duration of the heat treatments can be tuned to optimize the thermomechanical properties of the wire. An illustration was given here, where, by using higher temperature treatments, we switched from a superelastic to a ferroelasticity behavior with shape memory effect.

The thermomechanical properties of the entanglements were analyzed based on compression experiments. Although other types of mechanical loadings, such as tension or shear, should be considered to complete the experimental database, the results obtained here are interesting and promising. Entanglements processed in \textit{state 1} exhibit a remarkable superelastic behavior with negligible residual strains even after repeated cycles. This behavior is weakly strain-rate and temperature dependent. The nearly full reversion of the macroscopic strain observed upon unloading is explained by the unique association of superelastic wires together with their pronounced entanglement without strong physical bonds, allowing a reversible deformation and  rearrangement of the wire during the compression without plastic deformation or rupture. This was not observed with other metal foams, nor with other entangled metallic monofilaments. Combined with the relatively high stress levels and stiffnesses observed during compression (both of them could be tailored by increasing the fiber content), this material could be used as stiffer substitutes to standard elastomeric foams. The Young modulus also corresponds to values obtained for trabecular bones, so that combined with their porous mesostructure, these entanglements could also be relevant solutions for bone implants or scaffolds. 

The superelastic behavior is accompanied with a stress hysteresis, which increases with imposed strain and fiber content. Such a hysteresis is due to strain-induced rearrangements of the entanglements, to their associated friction forces, and to stress-induced forward and reverse martensitic transformations occurring in the wire. Likewise, the observed hysteretic behavior exhibits an interesting discrete memory, which needs to be clarified at the mesoscale and should also be taken into account in macroscale constitutive equations dedicated to model the mechanics of such materials. Lastly, we have shown that this pronounced hysteresis provides the entanglements with good damping properties, so that combined with their rather high stiffness, these entanglements can be viewed as suitable solutions for load bearing and damping structural components.

Switching from a superelastic to a ferroelastic state profoundly changes the mechanical behavior of the entanglements, leading to a marked degradation of both the stress-levels and strain recovery. This is directly related to an apparent plastic deformation of the wire, induced both by martensitic transformation and martensite reorientation. We have shown that the corresponding macroscale residual strain, even at levels as high as 0.16, can surprisingly be fully recovered upon stress-free heating. This full shape memory effect has never been observed with other NiTi porous materials. Once again, it is related to the unique association of NiTi wire with a high level of entanglement free of strong physical bonds.  

To optimise the architecture of these materials and to better understand both their remarkable macroscale properties and their underlying complex deformation meso-mechanisms, further and finer investigations are needed. In particular, performing similar experiments with 3D \textit{in situ} observations of the entanglement mesostructures, \textit{e.g.} by using X-ray microtomography \citep{Masse06,Latil11,Courtois12} and/or using discrete/finite element simulations at the fiber scale \cite{Rodney05,Durville05,Barbier09} would undoubtedly bring relevant elements towards this end. This work is in progress.

\vspace{0.5cm}

\textbf{Acknowledgements} - This work was performed within the ANR research programs ``3D discrete Analysis of micromechanisms of deformation in highly concentrated fiber suspensions'' (ANAFIB, ANR-09-JCJC-0030-01) and ``Architectured NiTi Materials'' (ANIM, N.2010 BLAN 90201). The 3SR Lab is part of the LabEx Tec21(Investissements
d'Avenir - grant agreement ANR-11-LABX-0030).

\linespread{1}
\small


%
\newpage


%

%

\end{document}